\documentclass[a4paper,12pt]{amsart}
\usepackage{amssymb}
\usepackage{amscd}

\newcommand{\qft}{quantum Fourier transform\;}

\newtheorem{theo}{Theorem}[section]
\newtheorem{lemm}{Lemma}[section]

\begin{document}
\title[error correction]{quantum error-correction codes on abelian groups}
\date{}
\author[M. Amini]{Massoud Amini}
\address{Department of Mathematics and Statistics\\University of Calgary\\ 2500 University Drive N.W., Calgary\\Alberta, Canada T2N 1N4\\
\linebreak
mamini@math.ucalgary.ca
\linebreak
Permanent address: Department of Mathematics\\Tarbiat Modarres University\\ B.O.Box 14115-175\\Tehran, Iran\\
amini@modares.ac.ir}
\keywords{quantum error correction, qunatum Fourier transform, quantum channel}
\subjclass{81P68}
\thanks{This research was done while I was visiting University of Calgary, I would like to thank the Deaprtment of Mathematics and Statistics in U of C for their support. I am also grateful to Professor Richard Cleve for his moral support.}
\begin{abstract}
We prove a general form of bit flip formula for the quantum Fourier transform
on finite abelian groups and use it to encode some general CSS codes on these groups.
\end{abstract}

\maketitle

\section{Introduction}

In classical public key cryptography the security of the cryptosystems are based on the difficulty of calculating certain functions. A famous example is the ASP cryptosystem which was based on the assumption that factoring large integers could not be done in polynomial time (on classical computers). The typical situation in these systems is that two parties (Bob and Alice) whish to communicate in secret. Instead of sharing a secrete key in advance (which confront us with the relatively difficult issue of secret key distribution), Bob  announces a public key which is used by Alice to encrypt a message, sent to Bob. The encryption is done in a clever way so that if a third party (Eve) wants to decrypt it a non feasible amount of calculation is needed. Bob, however, has a secret key of his own which enables him to do the encryption in real time.

Quantum cryptography has a different way of keeping things secret. The difficulty of some calculations is replaced by the impossibility of some calculations according to the laws of quantum mechanics. The first example of the quantum key distribution protocol was published in 1984 by Bennett and Brassard [BB] which is now called BB84 code. The security of this protocol is gauranteed by the impossibility of measuring the state of a quantum system in two conjugate bases simultaneously. A complete proof of security  against any possible attack (i.e. any combination of physical operations permitted by the laws of quantum mechanics) was given later [LC], [M], [BBMR]. A simple proof of this fact is proposed by Shor and Preskill in [SP]. They first showed the security of a modified Lo-Chau code which is a entanglement purification protocol and uses EPR pairs. Then they showed that it is equivalent to a quantum error correcting code, namely the CSS code introduced independently in [CS] and [S]. This later code was constructed on the vector space $\{0,1\}^n$ after the classical binary codes. Finally they reduced the CSS code to BB84. The basic idea of this final step was to avoid the quantum memory and reduce the encoding and decoding to classical computations.

The encoding part in the CSS protocol in [SP] was based on the following property of linear codes: If $\mathcal C$ is a linear code then the value of
$\frac{1}{|\mathcal C|}\sum_{y\in \mathcal C} (-1)^{x.y}$ is $1$ or $0$ when $x\in \mathcal C^\perp$ or $x\notin\mathcal C^\perp$, respectively. This is used to show that the Hadamard gate transforms the state
$$\frac{1}{\sqrt{|\mathcal C|}}\sum_{y\in \mathcal C} (-1)^{a.y}|y+b\rangle$$
to the state
$$\frac{(-1)^{a.b}}{\sqrt{|\mathcal C^\perp|}}\sum_{y\in \mathcal C^\perp} (-1)^{b.y}|y+a\rangle$$
In this paper we generalize this observation to the setting of arbitrary finite abelian groups (note that in linear coding theory $\{0,1\}^n$ is treated as a vector space, but it is also an abelian group). We show that for a finite abelian group $G$, a subgroup $H$, and elements $a,b\in G$, the quantum Fourier transform sends the state
$$\frac{1}{|H|}\sum_{z\in H} \overline{\chi_a(z)}|z+b\rangle$$
to the state
$$\frac{\chi_a(b)}{|H^\perp|}\sum_{z\in H^\perp} \chi_b(z)|z+a\rangle$$
where $\{\chi_x: x\in G\}$ is a Fourier basis for $G$ and $H^\perp=\{x\in G: \chi_x(y)=1\quad (y\in H)\}$. We use this to build a version of CSS code adapted to the group case. We show that the calculations of [SP] carries over and we can reduce this code to a generalized version of BB84 built on group $G$. The paper continues as follows. In section 2 we introduce the quantum Fourier transform on a finite abelian group $G$ and prove the above statement. In section 3 we discuss quantum error correction codes and introduce the CSS code on $G$. In the last section we above mentioned two protocols and show their equivalence.

\section{\qft}

Let $G$ be a finite abelian (additive) group. Let $\mathcal H=\mathbb CG$ be a Hilbert space with the
orthonormal basis $\{|x\rangle: x\in G\}$, called the {\it standard basis} of $\mathcal H$.
There is a natural action of $G$ on $\mathcal H$ by translation
$$x:|y\rangle\mapsto|x+y\rangle\quad (x,y\in G)$$
Note that $\mathbb CG$ is also an algebra under the convolution product
$$(\sum_{x\in G} c_x|x\rangle)*(\sum_{y\in G} d_x|y\rangle)=\sum_{z\in G} (\sum_{x+y=z} c_xd_y)|z\rangle$$
A {\it character} on $G$ is a nonzero group homomorphism
$\chi: G\to \mathbb T$, where $\mathbb T$ is the multiplicative group of the complex numbers of modulus 1.
The values $\chi(x)$ are $|G|$-th roots of unity. The set $\hat G$ of all characters on $G$ is an abelian group
with respect to the pointwise multiplication and is called the {\it dual group} of $G$. It is well known that $|\hat G|=|G|$ and so
we may index the elements of $\hat G$ by elements of $G$, and write $\hat G=\{\chi_x: x\in G\}$. Indeed in the finite group case we have,
$\hat G\simeq G$, so we may assume that $\chi_x\chi_y=\chi_{x+y}$ and $\chi_x(y)=\chi_y(x)$, for each $x,y\in G$, and $\chi_0\cong 1$. Also we have
the Schur's orthogonality relations
$$\frac{1}{|G|}\sum_{x\in G} \chi_y(x)\overline{\chi_z(x)}=\delta_{yz}\quad (y,z\in G).$$

For each $x\in G$ cosider the state
$$|\chi_x\rangle=\frac{1}{|G|}\sum_{y\in G} \overline{\chi_x(y)}|y\rangle,$$
then the above orthogonality relations imply that $\{|\chi_x\rangle: x\in G\}$ forms a orthonormal basis for $\mathcal H$, called the {\it Fourier
basis} of $\mathcal H$. This basis is translation invariant in the sense that
$$x|\chi_y\rangle=\chi_y(x)|\chi_y\rangle\quad (x,y\in G)$$
The {\it \qft} on $G$ is the unitary operator
$F_G:\mathcal H\to\mathcal H$ defined by
$$|x\rangle\mapsto \frac{1}{\sqrt{|G|}}\sum_{y\in G} \chi_x(y)|y\rangle\quad(x,y\in G)$$
Note that one can extend this map by linearity on $\mathcal H$ (see [J]). Two classical examples are $G=\mathbb Z_m$ where
$$\chi_k(\ell)=e^{2\pi ik/m}\quad k,\ell=0,\dots,m-1$$
and $G=\{0,1\}^n$ where
$$\chi_x(y)=(-1)^{x.y}\quad (x,y\in\{0,1\}^n)$$
in which $F_G$ is the usual discrete Fourier transform $DFT_m$ on $\mathbb Z_m$ and the Hadamard transform $H_n$, respectively.

Each element of $\hat G$ could be extended by linearity to a linear functional on $\mathbb CG$. This is indeed a multiplicative functional
with respect to the convolution product and $\hat G$ exhusts the set of all multiplicative linear functionals [R]. The well known Peter-Weil
theorem applied to the finite group $G$, tells us that $\hat G$ is an orthonormal basis for the linear dual space $(\mathbb CG)^{*}$.
In particular $(\mathbb CG)^{*}\simeq \mathbb C\hat G$. For each subset $H\subseteq G$, $\mathbb CH$ is a subspace of $\mathbb CG$,
generated by $\{|x\rangle: x\in H\}$. We put
$$H^\perp=\{x\in G: \chi_x(y)=1\quad(y\in H)\}$$
If $H$ is a subgroup of $G$ (we write $H\leq G$), then $H^\perp\simeq (G/H)\hat{}$ [R]. This notion goes in parallel with the notion of the orthogonal
complement $L^\perp$ for a subspace $L\leq \mathbb CG$. Of course $(\mathbb CH)^\perp$ and $\mathbb CH^\perp$ are not the same (even
the dimensions don't match).

\begin{lemm} If $H\leq G$ and $x\in G\backslash H^\perp$, then there is $K\leq H$ with $[H:K]=2$ and $x_0\in H$ of order two such that
$H=K\cup K+x_0$, $K\cap K+x_0=\emptyset$, and $\chi_x(x_0)=-1$.
\end{lemm}
{\bf Proof} Consider the subspace $L\leq \mathbb CH$ with $L^\perp=<\mathbb CH^\perp,x>$. Then $L$ has codimension $1$ in $\mathbb CH$,
so we can write $H=K\cup\{x_0\}$ for some $0\neq x_0\in H$ and $K\subseteq H$ with $L=\mathbb CK$ and $\mathbb CH=<L,x_0>$. Since
$0\in K$ so $x_0\in K+x_0$ and therefore $H\subseteq K\cup K+x_0$. But $H$ is a group, so $K\cup K+x_0\subseteq H$, that is
$H=K\cup K+x_0$. Now if $K\cap K+x_0\neq \emptyset$, then $x_0\in L$, which is not possible. To see that $K$ is a subgroup of
$H$ take $x,y\in K$, then $x-y\in H=K\cup K+x_0$, but $x-y\in K+x_0$ would imply that $x_0\in \mathbb CK=L$ which is again impossible, so
$x-y\in K$. Now $K$ has exactly two cosets in $H$, so $[H:K]=2$ and the group generated by $x_0$ is isomorphic to the quotient
group $H/K$ of order $2$, so $x_0$ has order $2$. In particular $\chi_x(x_0)=1$ or $-1$. But $x\in (\mathbb CK)^\perp$ so $\chi_x(k)=1$,
for each $k\in K$. Hence $\chi_x(x_0)\neq 1$ (otherwise $x\in H^\perp$), and so $\chi_x(x_0)=-1$.QED

\begin{lemm} For each $x\in G$ and $H\leq G$ we have
$$\sum_{y\in H} \chi_x(y)=\left\{\begin{array}{cc}
|H| &\mbox{if $x\in H^\perp$}\\
0 &\mbox{otherwise}
\end{array}
\right\}$$
\end{lemm}
{\bf Proof}
If $x\in H^\perp$ then
$$\sum_{y\in H} \chi_x(y)=\sum_{y\in H} 1=|H|$$
If $x\notin H^\perp$, then with the notation of the above lemma
\begin{align*}
\sum_{y\in H} \chi_x(y) &=\sum_{y\in K} \chi_x(y)+\sum_{y\in K+x_0} \chi_x(y)\\
&=\sum_{y\in K} \chi_x(y)+\sum_{y\in K} \chi_x(y+x_0)\\
&=\sum_{y\in K} (1+\chi_x(x_0))\chi_x(y)=0.QED
\end{align*}

\vspace{.3 cm}
For each $x,y\in G$ let $|x\rangle\langle y|$ be the rank one operator on $\mathcal H=\mathbb CG$ defined by
$$(|x\rangle\langle y|)|z\rangle=\langle y|z\rangle|x\rangle\quad(z\in G)$$
then one can decompose the \qft as a combination of rank one operators.

\begin{lemm}$F_G=\frac{1}{\sqrt{|G|}}\sum_{x,y\in G} \chi_x(y)|y\rangle\langle x|$.
\end{lemm}
{\bf Proof} If $F_G$ is defined by above formula, then for each $z\in G$
\begin{align*}
F_G|z\rangle&=\frac{1}{\sqrt{|G|}}\sum_{x,y\in G} \chi_x(y)|y\rangle\langle x|z\rangle=\frac{1}{\sqrt{|G|}}\sum_{x,y\in G} \chi_x(y)\delta_{xz}|y\rangle\\
&=\frac{1}{\sqrt{|G|}}\sum_{y\in G} \chi_x(y)|y\rangle.QED
\end{align*}

Now we are ready to prove the main result of this section.

\begin{theo} Let $a,b\in G$ and $H\leq G$ and consider the state

$$|\psi\rangle=\frac{1}{\sqrt{|H|}}\sum_{z\in H} \overline{\chi_a(z)}|z+b\rangle$$
then
$$F_G|\psi\rangle=\frac{\chi_a(b)}{\sqrt{|H^\perp|}}\sum_{z\in H^\perp} \chi_b(z)|z+a\rangle.$$
\end{theo}

{\bf Proof}
If we use the above lemma and the fact that
$$\chi_{z+b}(y)=\chi_z(y)\chi_b(y),\, \chi_z(y)=\chi_y(z)\quad (y,z\in G)$$
we have
\begin{align*}
F_G|\psi\rangle &=\frac{1}{\sqrt{|G||H|}}\sum_{x,y\in G} \chi_x(y)|y\rangle\langle x|\sum_{z\in H} \chi_{-a}(z)|z+b\rangle\\
&=\frac{1}{\sqrt{|G||H|}}\sum_{x,y\in G}\sum_{z\in H} \chi_x(y)\chi_{-a}(z)|y\rangle\langle x|z+b\rangle\\
&=\frac{1}{\sqrt{|G||H|}}\sum_{y\in G}\sum_{z\in H} \chi_{z+b}(y)\chi_{-a}(z)|y\rangle\\
&=\frac{1}{\sqrt{|G||H|}}\sum_{y\in G}\sum_{z\in H} \chi_{z}(y)\chi_b(y)\chi_{-a}(z)|y\rangle\\
&=\frac{1}{\sqrt{|G||H|}}\sum_{y\in G}\sum_{z\in H} \chi_{y}(z)\chi_b(y)\chi_{-a}(z)|y\rangle\\
&=\frac{1}{\sqrt{|G||H|}}\sum_{y\in G}\sum_{z\in H} \chi_{b}(y)\chi_{y-a}(z)|y\rangle\\
&=\frac{1}{\sqrt{|H|^\perp}}\big(\frac{1}{|H|}\sum_{z\in H}\chi_{y-a}(z)\big)\big(\sum_{y\in G} \chi_{b}(y)|y\rangle\big)\\
&=\frac{1}{\sqrt{|H|^\perp}}\sum_{y-a\in H^\perp} \chi_{b}(y)|y\rangle\\
&=\frac{1}{\sqrt{|H|^\perp}}\sum_{z\in H^\perp} \chi_{b}(z+a)|z+a\rangle\\
&=\frac{\chi_b(a)}{\sqrt{|H|^\perp}}\sum_{z\in H^\perp} \chi_{b}(z)|z+a\rangle\\
&=\frac{\chi_a(b)}{\sqrt{|H|^\perp}}\sum_{z\in H^\perp} \chi_{b}(z)|z+a\rangle.QED
\end{align*}

\section{quantum error correcting codes}

A {\it quantum channel} $Q$ is a trace preserving completely positive linear map
$$Q:\mathcal H_{in}\to\mathcal H_{out}$$
We can decompose $Q$ as
$$Q(\rho)=\sum_{i\in I} A_i\rho A_i^\dagger,$$
where $A_i$'s are {error operators} with $\sum_{i\in I}  A_i^\dagger A_i$ equal to the identity operator. In general $Q$ is not invertible, unless
restricted to a subspace. A subspace $\mathcal C\leq\mathcal H_{in}$ is called a {\it quantum error correcting code}(QECC) for $Q$
if there is a {\it decoding operator} $D$ such that
$$DQ|\psi\rangle\langle\psi|=\psi\rangle\langle\psi|\quad (\psi\in\mathcal C),$$
or equivalently
$$P_{\mathcal C}A_k^\dagger A_\ell P_{\mathcal C}=\alpha_{k\ell}P_{\mathcal C}\quad(k,\ell\in I),$$
for some constants $\alpha_{k\ell}$, where $P_{\mathcal C}$ is the projection operator onto $\mathcal C$ [K].

Now let $G$ be a finite abelian group and $G^n=G\times \dots \times G$ ($n$ copies). A subgroup $\mathcal C\leq G^n$ with
$k=|\mathcal C|$ is called a $[n,k]_G$ code. Elements of $G^n$ are words $x=(x_1,\dots,x_n)$ and the words in $\mathcal C$ are
called {\it codewords}. For $x,y\in G^n$, the {\it distance} $d(x,y)$ is the number of coordinates in which $x$ and $y$ differ.
The {\it weight} of a word $x$ is the number $wt(x)$ of its nonzero coordinates, where zero is the identity of $G$. A $[n,k]_G$
code with minimum distance $d$ is called a $[n,k,d]_G$ code. When $G=(\mathbb F_2,+)$, this is nothing but the classical binary
code $[n, log_2(k), d]$.

Suppose $\mathcal C_1$ and $\mathcal C_2$ are $[n,k_1]_G$ and $[n,k_2]_G$ codes with $\mathcal C_2\leq \mathcal C_1$ and $\mathcal C_1$
and $\mathcal C_2^\perp$ both correct $t$ errors. We define a quantum code $CSS_G(\mathcal C_1,\mathcal C_2)$ capable of correcting
errors on $t$ qubits. For a codeword $x\in \mathcal C_1$ put
$$|x+\mathcal C_2\rangle=\frac{1}{\sqrt{|\mathcal C_2|}}\sum_{y\in \mathcal C_2}|x+y\rangle$$
Note that $|x+\mathcal C_2\rangle$ only depends on the coset of $\mathcal C_1/\mathcal C_2$ to which $x+\mathcal C_2$ belongs.
Also $|x+\mathcal C_2\rangle$ is orthogonal to $|y+\mathcal C_2\rangle$, if $x$ and $y$ are representatives of different cosets
of $\mathcal C_2$. The quantum code $CSS_G(\mathcal C_1,\mathcal C_2)$ is defined on the vector space spanned by the states $|x+\mathcal C_2\rangle$
, where $x$ ranges in $\mathcal C_1$. In particular the dimension of $CSS_G(\mathcal C_1,\mathcal C_2)$ is $|\mathcal C_1|/|\mathcal C_2|$.

Suppose that a bit flip and a phase flip errors have occured. These are described by two "$n$ bit"  vectors $e_1,e_2\in G^n$.
If $|\psi\rangle=|x+\mathcal C_2\rangle$ is the original state, then the corrupted state would be
$$|\psi_1\rangle= \frac{1}{\sqrt{|\mathcal C_2|}}\sum_{y\in \mathcal C_2}\chi_{e_2}(x+y)|x+y+e_1\rangle$$
as in the binary case, the encoding process starts with introducing a {\it ancilla} (of sufficient length) initially in the all
zero state $|0\rangle$. We apply the parity matrix $H_1$ for the code $\mathcal C_1$ taking $|x+y+e_1\rangle|0\rangle$  to
$$|x+y+e_1\rangle|H_1(x+y+e_1\rangle=|x+y+e_1\rangle|H_1e_1\rangle$$
where the above equality follows from the fact that $x+y\in \mathcal C_1$, and so $H_1(x+y)=0$. The effect of this operation on
$|\psi_1\rangle|0\rangle$ is
$$\frac{1}{\sqrt{|\mathcal C_2|}}\sum_{y\in \mathcal C_2}\chi_{e_2}(x+y)|x+y+e_1\rangle|H_1e_1\rangle$$
Now error detection for the bit flip error is simply done by measuring the ancilla. This gives us $H_1e_1$, from which we can
infer $e_1$, since $\mathcal C_1$ can correct up to $t$ errors. The result of discarding the ancilla is the state
$$|\psi_2\rangle= \frac{1}{\sqrt{|\mathcal C_2|}}\sum_{y\in \mathcal C_2}\chi_{e_2}(x+y)|x+y+e_1\rangle$$
Next applying the $U_{e_1}^\dagger: |z\rangle \mapsto |z-e_1\rangle$ unitary gate to this state, we obtain
$$|\psi_3\rangle= \frac{1}{\sqrt{|\mathcal C_2|}}\sum_{y\in \mathcal C_2}\chi_{e_2}(x+y)|x+y\rangle$$
The next step is applying the \qft $F_{G^n}=F_G\otimes\dots\otimes F_G$ ($n$ times) to $|\psi_3\rangle$. Using Theorem 2.1 (applied
to $G^n$ with $H=\mathcal C_2$, $a=-e_2$, and $b=x$) we get
\begin{align*}
|\psi_4\rangle &=F_{G^n}|\psi_3\rangle= \chi_{e_2}(x)F_{G^n}\big(\frac{1}{\sqrt{|\mathcal C_2|}}\sum_{y\in \mathcal C_2}\chi_{e_2}(y)|y+x\rangle\big)\\
&=\frac{\chi_{e_2}(x)\chi_{-e_2}(x)}{\sqrt{|\mathcal C_2^\perp|}}\sum_{y\in \mathcal C_2^\perp}\chi_{x}(y)|y-e_2\rangle\\
&=\frac{1}{\sqrt{|\mathcal C_2^\perp|}}\sum_{y\in \mathcal C_2^\perp}\chi_{x}(y)|y-e_2\rangle
\end{align*}
As for the error detection for the bit flip, we introduce an ancilla and apply the parity matrix $H_2$ for $\mathcal C_2^\perp$ to
obtain $H_2(-e_2)$, and correct the phase flip error (now showing up as a bit flip error), obtaining the state
$$|\psi_5\rangle=\frac{1}{\sqrt{|\mathcal C_2^\perp|}}\sum_{y\in \mathcal C_2^\perp}\chi_{x}(y)|y\rangle$$
Again applying $F_{G^n}$ and using Theorem 2.1 (with $H=\mathcal C_2^\perp$, $a=-x$, and $b=0$) we get
\begin{align*}
|\psi_6\rangle &=F_{G^n}|\psi_5\rangle=\frac{\chi_{-x}(0)}{\sqrt{|\mathcal C_2|}}\sum_{y\in \mathcal C_2}\chi_{0}(y)|y-x\rangle\\
&=\frac{1}{\sqrt{|\mathcal C_2|}}\sum_{y\in \mathcal C_2}|y-x\rangle
\end{align*}
Finally, applying the $U_{x}^2: |z\rangle \mapsto |z+x+x\rangle$ unitary gate to this state, we get back our original state
$$|\psi\rangle=\frac{1}{\sqrt{|\mathcal C_2|}}\sum_{y\in \mathcal C_2}|y+x\rangle$$
with a slight modification of the above proof, we have

\begin{theo} Suppose $\mathcal C_1$ and $\mathcal C_2$ are $[n,k_1,d_1]_G$ and $[n,k_2,d_2]_G$ codes with $\mathcal C_2\leq \mathcal C_1$, let
$V=\{v_1,\dots,v_k\}$ be the set of representatives of the quotient group $\mathcal C_1/\mathcal C_2$, then the $k=\frac{k_1}{k_2}$ mutually
orthogonal states
$$|\psi_i\rangle=\frac{1}{\sqrt{|\mathcal C_2|}}\sum_{y\in \mathcal C_2}|y+v_i\rangle$$
are a basis for a quantum error correction code $\mathcal C\leq \mathcal H^{\otimes^n}$, where $\mathcal H=\mathbb CG$ is the
group algebra of $G$. The code can simultaneously correct at least $\lfloor \frac{d_1-1}{2}\rfloor$ spin flip errors and
$\lfloor \frac{d_2-1}{2}\rfloor$ phase flip errors. Its minimum distance is $d\geq min{d_1,d_2}$. We denote this QECC by
$CSS_G(\mathcal C_1,\mathcal C_2)$ or $[[n,k,d]]_G$.
\end{theo}

\section{a quantum error correction protocol}
In this section we use a version of the quantum error correction code $CSS_G(\mathcal C_1,\mathcal C_2)$ to write a quantum error correction protocol similar
to the protocol introduced in [CS] (for the case $G=\mathbb F_2$). Let $\mathcal C_1$ and $\mathcal C_2$ be as in the Theorem  3.1, for
each $x\in\mathcal C_1^\perp$ and $z\in \mathcal C_2$ consider the quantum error correction code $CSS_G^{z,x}(\mathcal C_1,\mathcal C_2)$
with codeword states
$$|\psi_{v,z,x}\rangle=\frac{1}{\sqrt{|\mathcal C_2|}}\sum_{w\in \mathcal C_2}\chi_z(w)|v+w+x\rangle$$
where $v$ ranges over the representatives of the $|\mathcal C_1|/|\mathcal C_2|$ cosets of $\mathcal C_2$ in $\mathcal C_1$ (we use the
notation $[v]$ as an abbriviation for the coset $v+C_2$. Note that the number of these states is
$$|\mathcal C_1|/|\mathcal C_2|.|\mathcal C_2|.|\mathcal C_1^\perp|=|G^n|=|G|^n$$
We show that these states are mutually orthogonal, and therefore form a basis for an $|G|^n$-dimensional vector space.

\begin{lemm}
$\sum_{z\in C_2} |\psi_{v,z,x}\rangle\langle\psi_{v,z,x}|=\sum_{w\in C_2} |v+w+x\rangle\langle v+w+x|$
\end{lemm}
{\bf Proof}
Using Lemma 2.2 applied to $G^n$ (with $H=\mathcal C_2$ and $x=w_1-w_2$) we have
\begin{align*}
\sum_{z\in C_2} |\psi_{v,z,x}\rangle\langle\psi_{v,z,x}|&=\frac{1}{|\mathcal C_2|}\sum_{z\in C_2} \sum_{w_1,w_2\in \mathcal C_2}\chi_z(w_1-w_2)|v+w_1+x\rangle\langle v+w_2+x|\\
&=\sum_{w_1,w_2\in \mathcal C_2}\big(\frac{1}{|\mathcal C_2|}\sum_{z\in C_2} \chi_{w_1-w_2}(z)\big)|v+w_1+x\rangle\langle v+w_2+x|\\
&=\sum_{w_1,w_2\in \mathcal C_2}\delta_{w_1,w_2}|v+w_1+x\rangle\langle v+w_2+x|\\
&=\sum_{w\in C_2} |v+w+x\rangle\langle v+w+x|.QED
\end{align*}

Let us use the abbreviation $\sum_{v,z,x}$ to denote the summation over all $[v]\in \mathcal C_1/\mathcal C_2$, $z\in \mathcal C_2$,
and $x\in \mathcal C_1^\perp$.

\begin{lemm}
$\sum_{v,z,x} |\psi_{v,z,x}\rangle\langle\psi_{v,z,x}|=I$, the identity operator on $\mathbb CG^n$.
\end{lemm}
By above lemma
$$\sum_{v,z,x} |\psi_{v,z,x}\rangle\langle\psi_{v,z,x}|=\sum_{v,x}\sum_{w\in C_2} |v+w+x\rangle\langle v+w+x|$$
but each $y\in G^n$ has a unique decomposition $y=v+w+x$, for some $[v]\in \mathcal C_1/\mathcal C_2$, $w\in \mathcal C_2$,
and $x\in \mathcal C_1^\perp$. Therefore the last sum is the same as
$$\sum_{y\in G^n} |y\rangle\langle y|=I.QED$$

A similar argument proves

\begin{lemm}
$\sum_{v,z,x} |\psi_{v,z,x}\rangle|\psi_{v,z,x}\rangle=\sum_{y\in G^n} |y\rangle|y\rangle$.
\end{lemm}

Now we are ready to present our quantum error correction protocol. It is based on the modified Lo-Chau protocol[LC] and follows the presentation of a similar construction as reported in [NC]. It uses our quantum error correction code to perform entanglement distillation. The basic difference here is the meaning of a "qubit". For us a qubit is a basis element of $\mathcal H=\mathbb CG$, namely a state of the form $|t\rangle$, where $t\in G$ (bit has a similar meaning). Also let us remind that the standard basis of $\mathcal H$ is $\{|t\rangle : t\in G\}$. So for the given finite abelian group $G$, we have the following protocol.

\begin{center}
{\bf QKD protocol: $CSS_G$ codes}
\end{center}

\vspace{.2 cm}
{\bf 1:} Alice creates $n$ random check bits, a random $m$ bit key $k$, and two random $n$ bit strings $x$ and $z$. She encodes $|k\rangle$ in the code $CSS_G^{z,x}(\mathcal C_1,\mathcal C_2)$. She also encodes $n$ qubits according to the check bits.

\vspace{.2 cm}
{\bf 2:} Alice randomely chooses $n$ positions (out of $2n$) and puts the check qubits in these positions and the encoded qubits in the remaining positions.

\vspace{.2 cm}
{\bf 3:} Alice selects a random $2n$ bit string $b$ and performs a Fourier transform $F_G$ on each qubit for which $b$ is not $0$ ($0$ is the identity of $G$).

\vspace{.2 cm}
{\bf 4:} Alice sends the resulting qubits to Bob.

\vspace{.2 cm}
{\bf 5:} Bob receives the qubits and publicly announces this fact.

\vspace{.2 cm}
{\bf 6:} Alice announces $b$, $z$, $x$, and which $n$ qubits are to provide check bits.

\vspace{.2 cm}
{\bf 7:} Bob performs the Fourier transform on the qubits where $b$ is not $0$.

\vspace{.2 cm}
{\bf 8:} Bob measures the $n$ check qubits in the standard basis, and publicly shares the results with Alice. If more than $t$ of these disagree, they abort the protocol.

\vspace{.2 cm}
{\bf 9:}  Bob decodes the remaining $n$ qubits from $CSS_G^{z,x}(\mathcal C_1,\mathcal C_2)$.

\vspace{.2 cm}
{\bf 10:}  Bob measures his qubits to obtain the shared secret key $k$.

\vspace{.2 cm}
A series of remarks are in order. We have emplyed $CSS_G(\mathcal C_1,\mathcal C_2)$ code, which we assumed to encode $m$ qubits in $n$ qubits and correct up to $t$ errors. The Alice's $n$ EPR pair state may be written as the equal states given in Lemma 4.3. Note the lables are separated to indicate the qubits Alice keeps, and the ones she sends to Bob. If Alice wants to measure her remaining qubits according to the check matrix for $CSS_G(\mathcal C_1,\mathcal C_2)$, she obtains random values for $x$ and $z$, and if she wants to measure the $m$ EPR pair in the standard basis, she obtains a random choice of $v$. Then the remaining $n$ qubits are left in the state $|\psi_{v,z,x}\rangle$, which is the codeword for $v$ in $CSS_G^{z,x}(\mathcal C_1,\mathcal C_2)$ and is the encoded version of the state $k\rangle$.

Following [SP], one may do the following modifications in the protocol. Bob measures his qubits in the standard basis (which is e version of the $Z$ basis in the binary case) after decoding so the phase correction sent as $z$ by Alice is irrelevant. Therefore, instead of decoding and then measuring, Bob can immidiately measure to obtain $v+w+x$ (up to some error), then decode (classically ) as follows. He can subtract the annonced value of $x$ and coorect the result to a codeword in $\mathcal C_1$, which would be $v+w$ if the distance of the code is not exceeded. Then the key $k$ is the coset $v+w+\mathcal C_2$ in $\mathcal C_1$. Now as Alice need not reveal $z$, she is effectively sending a mixed state averaged over random values of $z$, which by Lemma 4.1 is
$$\frac{1}{|G|^n}\sum_{w\in C_2} |v+w+x\rangle\langle v+w+x|$$
To create this state, Alice only needs to choose $w\in \mathcal C_2$ randomely and construct $|v+w+x\rangle$ with her random values of $x$ and $k$.  Also if Alice happens to choose $v\in \mathcal C_1$ (rather than $[v]\in \mathcal C_1/\mathcal C_2$), then $w$ is unnecessary. In this case, Alice may choose $x$ at random, send $|x\rangle$ so that Bob receives and measures $x$ (with some error), then Alice sends $x-v$, which is subtracted by Bob to obtain $v$ (with some error). This leaves no difference between the random check bits and the code bits. Finally to avoid the performance of the Fourier transform by Alice, she can encode her qubits in the standard basis $\{|t\rangle : t\in G\}$ or the Fourier basis $\{|\chi_t\rangle : t\in G\}$, according to the bits of $b$, where

$$|\chi_t\rangle =\frac{1}{|G|}\sum_{s\in G} \overline{\chi_t(s)}|s\rangle$$

Then Bob could measure the received qubits randomely in the standard or Fourier bases. When Alice subsequently annonces $b$, they can keep only those bits for which their bases were the same. As they are most likely to discard half of their bits, they should start with a little more than twice the number of original random bits. This way Alice can delay her choice of check bits until after discarding. This allows us to avoid the use of quantum memory and perform the encoding and decoding classically. Summing up we have the following version of {\bf BB84}, adapted to the group $G$.

\begin{center}
{\bf QKD protocol: $BB84_G$ codes}
\end{center}

\vspace{.2 cm}
{\bf 1:} Alice creates $(4+\delta)n$ random bits.

\vspace{.2 cm}
{\bf 2:} Alice creates for each bit a qubit in the standard or Fourier basis, according to a random bit string $b$ (uses standard basis if at bits for which $b$ is $0$, and the Fourier basis otherwise).

\vspace{.2 cm}
{\bf 3:} Alice sends the resulting qubits to Bob.

\vspace{.2 cm}
{\bf 4:} Alice chooses a random $v\in \mathcal C_1$.

\vspace{.2 cm}
{\bf 5:} Bob receives the qubits, publicly announces this fact, and measures each in the standard or Fourier basis at random.

\vspace{.2 cm}
{\bf 6:} Alice announces $b$.

\vspace{.2 cm}
{\bf 7:} Alice and Bob discard those bits Bob measured in a basis other than the one instructed by $b$. With high probability, there are at least $2n$ bits left (if not abort the protocol). Alice decides randomely on a set of $2n$ bits to continue to use, randomely selects $n$ of these to be check bits, and announces the selecrtion.

\vspace{.2 cm}
{\bf 8:} Alice and Bob publicly compare their check bits. If more than $t$ of these disagree, they abort the protocol. Alice is left with the $n$ bit string $x$, and Bob with $x+\varepsilon$.

\vspace{.2 cm}
{\bf 9:} Alice annonces $x-v$. Bob subtracts this from his result, correcting it with code $\mathcal C_1$ to obtain $v$.

\vspace{.2 cm}
{\bf 10:}  Alice and Bob compute the coset $v+\mathcal C_2$ in $\mathcal C_1/\mathcal C_2$ to obtain the key $k$.


\begin{thebibliography}{99}

\bibitem
[BB]] C. H. Bennet, G. Brassard, Quantum cryptography: Public key distribution and coin tossing, in Proceedings of IEEE International Conference on Computers, Systems and Signal rocessing, Banglore, India, 1984, pp. 175-179. Also available as the IBM Technical Disclosure Bulletine 28 (1985), 3153-3163.

\bibitem
[BBBMR]] E. Biham, M. Boyer, P.O. Boykin, T. Mor, V. Roychowdhury, A proof of the security of quantum key distribution, in Proceedings of the Thirty Second Annual ACM Symposium on Theory of Computation, 2000, pp. 715-724. Also available online at arXive: quant-ph/9912053.

\bibitem
[CS]] A.R. Calderbank , P. Shor. Good quantum error correcting codes exist, Phys. Rev. A 54 (1996), 1098-1105, also available online at arXive: quant-ph/9512032.

\bibitem
[K]] K. Kraus, States, effects, and operations, Lect. Notes Phys., vol. 190, Speringer-Verlag, Berlin, 1983.

\bibitem
[J]] R. Jozsa , Quantum algorithms and the Fourier transform, quantum coherence and decoherence, Roy. Soc. Lond. Proc. Series  A, 454 (1998), no. 1969, 323-332,
also available online at arXiv:quant-ph/97033.

\bibitem
[LC]] H.K. Lo, H. F. Chau, Unconditional security of quantum key distribution over arbitrary long distances, Science 283 (1999) 2050-2056. Also available online at arXive: quant-ph/9803006.

\bibitem
[LC]] D. Meyers, Unconditional security in quantum cryptography, J. Assoc. Computing Machinery 48  (3) (2001), 351-406. Also available online at arXive:quant-ph/9802025.

\bibitem
[NC]] M.A. Nielsen, I. L. Chuang, Quantum computation and quantum information, Cambridge university press, Cambridge, 2000.

\bibitem
[R]] W. Rudin, Fourier analysis on groups, John Wiley and Sons, New York, 1990.

\bibitem
[SP]] P.W. Shor, J. Priskill, Simple proof of security of the BB84 quantum key distribution protocol, also available online at arXive quant-ph/0003004.

\bibitem
[S]] A. M. Steane, Multiple particle interference and error correction, Proc. R. Soc. London, series A, 452 (1996), 2551-2577.

\end{thebibliography}
\end{document}